\documentstyle{article}
\title {State property systems and closure spaces: \\ a study of categorical
equivalence\footnote{Published as: Aerts, D, Colebunders, E., Van der Voorde, A. and Van Steirteghem,
B., 1999, ``State property systems and closure spaces: a study of categorical
equivalence", {\it International Journal of Theoretical Physics}, {\bf 38}, 259.}}
\author {Diederik Aerts, Eva Colebunders, Ann Van der Voorde and \\ Bart Van Steirteghem}
\date {}

\newtheorem{theorem}{Theorem}
\newtheorem{definition}{Definition}

\newtheorem{proposition}{Proposition}

\newcommand{\qed}{\mbox{} \hfill $\Box$} 

\begin{document}

\maketitle
\centerline{FUND and TOPO,}
\centerline{Department of Mathematics, Brussels Free University,} \centerline {Pleinlaan 2, B-1050 Brussels,
Belgium} 
\centerline{CLEA,}
\centerline{Brussels Free University,}
\centerline{Krijgskundestraat 33, 1160 Brussels, Belgium}
\centerline{e-mails: diraerts@vub.ac.be, evacoleb@vub.ac.be,} \centerline{avdvoord@vub.ac.be,
bvsteirt@vub.ac.be} \bigskip

\medskip

\begin{abstract}
\noindent We show that the natural mathematical structure to describe a physical entity by means of its states and its properties within the Geneva-Brussels approach is that of a state property system. We prove that the category of state property systems (and morphisms), $\bf SP$, is equivalent to the
category of closure spaces (and continuous maps), $\bf Cls$. We show the equivalence
of the `state determination axiom' for state property systems with the `T$_{_0}$
separation axiom' for closure spaces. We also prove that the category $\bf SP_{_0}$
of state determined state property systems is equivalent to the category $\bf L_{_0}$ of based complete lattices. In this sense the equivalence of $\bf SP$
and $\bf Cls$ generalizes the equivalence of $\bf Cls_{_0}$ (T$_{_0}$ closure spaces) and
$\bf L_{_0}$, proven in (Ern\'{e} 1984). \end{abstract}

\section{Introduction}

Constantin Piron started the elaboration of a realistic axiomatic theory for the
foundations of quantum mechanics in Geneva and the first presentation of this
approach can be found in (Piron 1976). Apart from an axiomatic scheme presented
in (Piron 1976) - going back and founded on his celebrated representation
theorem (Piron 1964) - a first step of `operational' foundation was exposed in
(Piron 1976) by introducing the concept of `test' for a property. One of the
authors of the present paper (D. Aerts) studied the problem of the description
of `separated physical entities' within this approach. Making use extensively of
the `operational' idea presented in (Piron 1976), Aerts elaborated the
`operational' aspects of the theory, introducing a powerful `calculus of tests'
(Aerts 1981, 1982). In this way the theory grew to a complete realistic and
operational theory, and the `operational' part showed to be very well suited to
tackle `physical' problems, like the one of the description of separated
entities (Aerts 1981, 1982) and the filtering out of the classical part of an
entity (Aerts 1983). At this moment the theory is further elaborated in Geneva
and in Brussels and therefore we shall refer to it as the Geneva-Brussels
approach to the Foundations of Physics. It is a `realistic' and `operational'
theory, where a physical entity is described by means of its states and
properties, and the properties are `operationally' introduced as `testable
properties'. 

The foundation material
of the approach can be found in (Piron 1976, 1989, 1990, Aerts 1981, 1982, 1983)
and we will therefore refer to these writings as the foundation papers.
Meanwhile different problems have been investigated within the approach and
connections with other approaches to the Foundations of Physics have been
studied (Aerts 1981, 1982, 1983, 1984, 1985, 1994, 1998, Aerts, Coecke, Durt and
Valckenborgh 1997, Aerts and Valckenborgh 1998, Cattaneo et al. 1988, Cattaneo
and Nistico 1991, 1992, 1993, Daniel 1982, d'Emma 1980, Foulis et al. 1983,
Foulis and Randall 1984, Giovannini and Piron 1979, Gisin 1981, Jauch and Piron
1965, Ludwig and Neumann 1981, Moore 1995, Piron 1964, 1969, 1976, 1989, 1990,
Randall and Foulis 1983).

Although the foundational setting for the Geneva-Brussels approach was elaborated
in (Piron 1976, 1989, 1990, Aerts 1981, 1982, 1983), the basic mathematical
structure of the approach independent of the physical content had not yet been
properly identified. This was started in (Aerts 1998) within a more general
setting and the resulting mathematical structure has been called there a `state
property system' (see Aerts 1998 and section \ref{sec:phys} of this article). It
is shown - as we will do again in section~\ref{sec:phys} of this paper - that
the mathematical structure of a state property system, once the objects and
morphisms are given their physical meaning, manages to represent all the
subtleties of the approach. This has the enormous advantage that theorems can
now be proven within the approach without using the 'physical interpretation'
during the proofs of the theorems: an indispensable step for a real
formalization of the theory.

Moreover it is proven that state property systems and their morphisms are in
natural correspondence with closure spaces and continuous maps (Aerts 1998). In
the present paper we want to investigate this correspondence in detail: we show
that the category of state property systems and its morphisms, which we call
${\bf SP}$, is equivalent to the category ${\bf Cls}$ of closure spaces and continuous maps. This gives us a ``lattice representation'' for {\em all}
closure spaces. It generalizes older (well-known) lattice representations where
the closure spaces were (at least) T$_{_0}$ (Ern\'{e} 1984): if we restrict ourselves to T$_{_0}$ closure spaces, we recover the categorical equivalence between `based complete lattices' and T$_{_0}$ closure spaces, given in
(Ern\'{e} 1984) (sections
\ref{sec:T0} and
\ref{sec:embstate} of this paper).

The mathematical structure of a state property system that we will present in
this paper appears as the formalization of a state property entity within the
Geneva-Brussels approach. We want to remark however that it appears also as a
fundamental mathematical structure in other situations where states and
properties of physical entities are formalized (e.g.\ the situation presented in
(Aerts 1998) of a experiment state outcome entity with one experiment).

We remark that the description of a physical entity by means of its states and properties that we use in this article differs from the one in the founding papers (Piron 1976, 1989, 1990, Aerts 1981, 1982, 1983) in two aspects:

\smallskip
\noindent
(1) We make an explicit distinction between the properties and the states. In the founding papers a state of an entity is represented by the set of all actual properties, making it impossible to introduce the distinction as we will do it here. The distinction between states and properties was introduced in (Aerts 1994), where it was shown that a condition of `state determination' for an entity reduces this more general situation to the earlier one. It was also shown that the `state determination' condition is equivalent to the T$_{_0}$ separation axiom of the corresponding closure space. In (Aerts 1994) the categorical equivalence between the description of an entity by means of states and properties and the representation in the corresponding closure space was not yet elaborated: this will be the main subject of the present paper.

\smallskip
\noindent
(2) We explicitly distinguish between the physical content and the mathematical form of the theory. This was not done systematically in the founding papers and neither in (Aerts 1994). In (Aerts
1998), where such a systematic distinction between the physical and the mathematical is introduced for a more general theory also containing experiments and outcomes, the fruitfulness of this distinction became clear. It leads to the definition of the `mathematical' concept of a `state property system', representing the states and the properties of a general physical entity. This concept will be the central mathematical ``object'' in the present paper. We will show in a forthcoming paper in which way the categories formulated in the present paper are connected to the categories presented in (Moore 1995).

\section{The description of an entity by means of its states and testable properties \label{sec:phys}}

Let us consider an entity $S$. The entity $S$ is at every moment in a definite state $p$, and let us call $\Sigma$ the well defined set of considered states of the entity $S$.

If we have in mind a certain
property $a$ that the entity might have and if this property is testable, we can construct a test $\alpha$ for $a$. Such a test, also sometimes called `question' or `experimental project' in (Piron 1976, 1989, 1990, Aerts 1981, 1982, 1983), consists of an experiment that can be performed on the entity. If the experiment gives us the expected outcome, we will say that the answer to the test is `yes'. If the experiment does not give us the expected outcome, we will say that the answer to the test is `no'. Hence to define a test one has to give (1) the measuring apparatus used to perform the experiment connected to the test, (2) the manual of operation of the apparatus, and (3) a rule that allows us to interpret the results in terms of `yes' and `no'. Let us denote a well defined set of tests for an entity $S$ by $Q$.

We will say that a test $\alpha$ of the entity $S$ in a state $p$ is `true', and the corresponding property $a$ is `actual', if we can predict with certainty that the answer `yes' would come out if we were to perform the test.

The way that we introduced the
concepts of state, property, test, `true' test, and `actual' property, is till now equivalent with the way they are introduced in the founding papers. As
we have remarked we want to make an explicit distinction between the physical content of the theory and its mathematical form. That is the reason we introduce some additional concepts now. 

For a state
$p$ we consider the set $\eta(p)$ of all tests $\alpha \in Q$ that are `true' if the entity is in state $p$. Let us give now a formal definition of an entity described by its states and its set of testable properties. 

\begin{definition} [state test entity]
A state test entity $S$ is defined by a set $\Sigma$ (the set of states), a set $Q$ (the set of tests) and a function: \begin{equation}
\eta: \Sigma \rightarrow {\cal P}(Q): p \mapsto \eta(p) \end{equation}
where $\eta(p)$ is, by definition the set of tests which are `true' if the entity $S$
is in state
$p$. We call
$\eta$ the state test function.
Hence, for a test $\alpha \in Q$ and a
state $p \in \Sigma$ we have:
\begin{equation}
\alpha \ {\rm is \ true\ if}\ S\ {\rm is\ in\ state\ } p \Leftrightarrow \alpha \in \eta(p)
\end{equation}
We denote a state test entity $S$ as $S(\Sigma, Q, \eta)$. \end{definition}
If the situation is such that whenever the entity $S$ is in a state such that the test $\alpha$ is `true' then also the test $\beta$ is `true', we say that $\alpha$ `implies' $\beta$ (or $\alpha$ `is stronger' than $\beta$) and we denote $\alpha < \beta$. Let us now formally introduce the `implication' on the set of tests of a state test entity.
\begin{definition} [test implication]
Consider a state test entity $S(\Sigma, Q, \eta)$. For $\alpha, \beta \in Q$ we define:
\begin{equation}
\alpha < \beta \Leftrightarrow {\rm if\ for\ } p \in \Sigma \ {\rm we\ have\ } \alpha \in \eta(p)\ {\rm then\ } \beta \in \eta(p) \end{equation}
and we say that $\alpha$ 'implies' $\beta$ and call this relation the 'test implication'.
\end{definition}
We have a natural implication on the states that was not identified properly in the founding papers. If for two states $p, q \in \Sigma$ the set
$\eta(p)$ of all tests that are 'true' if the entity is in state $p$, includes the set
$\eta(q)$ of all tests that are 'true' if the entity is in state $q$, we say that
$p$ implies
$q$ (or $p$ is stronger than $q$) and we write $p < q$.

\begin{definition} [state implication]
Consider a state test entity $S(\Sigma, Q, \eta)$. For $p, q \in \Sigma$ we define:
\begin{equation}
p < q \Leftrightarrow \eta(q) \subset \eta(p) \end{equation}
and we say that $p$ `implies' $q$ and call this relation the `state implication'.
\end{definition}

\begin{proposition}
Consider a state test entity $S(\Sigma, Q, \eta)$. The implications on $Q$ and $\Sigma$ are pre-order relations. \end{proposition}
For a non-empty family of tests $(\alpha_i)_i$ we operationally introduce a product test
$\Pi_i\alpha_i$, like in the founding papers. It consists of choosing one of the
$\alpha_i$, performing this chosen test, and considering the outcome obtained as the outcome of
$\Pi_i\alpha_i$. We clearly have that $\Pi_i\alpha_i$ is true if and only if
$\alpha_i$ is true for each $i$. This means that $\Pi_i\alpha_i \in \eta(p)$ if and only if $\alpha_i \in \eta(p)\ \forall\ i$. Let us introduce the concept of `product test' formally. 

\begin{definition} [product test]
Consider a state test entity $S(\Sigma, Q, \eta)$ and a set $(\alpha_i)_i \in Q$ of tests. A product test $\Pi_i\alpha_i$ is a test such that:
\begin{equation}
\Pi_i\alpha_i \in \eta(p) \Leftrightarrow \alpha_i \in \eta(p)\ \forall\ i \label{prod}
\end{equation}
\end{definition}
We remark that the notation $\Pi_i\alpha_i$ for a product test is somewhat misleading. Indeed, in general a product test $\Pi_i\alpha_i$ does not have to be a test `formed' by the $\alpha_i$, as it is the case in the physical example that inspired the formal definition. It is just a test that satisfying the requirement expressed in formula (\ref{prod}). We remark that this mathematical definition of a product test makes sense for an empty family. In that case, it becomes a test which is always true. This type of test will be formally defined a
little further.

\begin{proposition} \label{compl}
Suppose that we have a state test entity $S(\Sigma, Q, \eta)$. If an arbitrary family of tests $(\alpha_i)_i \in Q$ has a product test $\Pi_i\alpha_i
\in Q$, then this product test is an infimum of the $(\alpha_i)_i$ in $Q,<$. \end{proposition}
Proof: We clearly have $\Pi_i\alpha_i < \alpha_j\ \forall\ j$. Suppose that $\beta < \alpha_i\ \forall\ i$, and consider a state $p \in \Sigma$ such that $\beta \in \eta(p)$. Then we have $\alpha_i \in \eta(p)\ \forall\ i$. As a consequence we have $\Pi_i\alpha_i \in \eta(p)$. This shows that $\beta < \Pi_i\alpha_i$. \qed

\bigskip
\noindent
We can define the following test: we do anything that we want with the entity and just give the answer `yes'. Clearly this test is always `true'. We can also introduce the following test: we do anything with the entity and just give the answer `yes' or `no' as we wish. Clearly this test is never `true'. Let us define these special types of tests formally.

\begin{definition} [unit and zero tests] Consider a state test entity $S(\Sigma, Q, \eta)$. We say that a test $\tau$ is a unit test if $\tau \in \eta(p)\ \forall\ p \in \Sigma$. We say that a test $\delta$ is a zero test if $\delta \notin \eta(p)\ \forall\ p \in \Sigma$.
\end{definition}

\begin{proposition} \label{unital}
Consider a state test entity $S(\Sigma, Q, \eta)$. If $\tau$ is a unit test we have for $\alpha \in Q$ that $\alpha < \tau$. If $\delta$ is a zero test we have for $\alpha \in Q$ that $\delta < \alpha$. \end{proposition}
Proof: Easy verification. \qed

\bigskip
\noindent
In the founding papers it is supposed that for each non-empty family of tests $(\alpha_i)_i \in Q$ there is a product test $\Pi_i\alpha_i \in Q$. And it is also supposed that there exists a unit test $\tau \in Q$ and a zero test $\delta \in Q$. Let us introduce these requirements on the formal level.

\begin{definition} [unital product entity] Suppose that we have a state test entity $S(\Sigma, Q, \eta)$. We say
that the entity is a `unital product' entity if $Q$ contains a unit test $\tau$, a
zero test $\delta$ and if for each family $(\alpha_i)_i \in Q$ there is a product
test
$\Pi_i\alpha_i
\in Q$.
\end{definition}
We remark that, since a product test of the empty family is an always true test,
demanding the existence of a unit test is in fact redundant. 

\begin{proposition}
Consider a unital product entity $S(\Sigma, Q, \eta)$. Then
for each set
$(\alpha_i)_i
\in Q$ of tests there exists an infimum and a supremum for the pre-order relation on Q. Further we have, for each unit test $\alpha$ and zero test $\beta$, and for a set of tests $(\alpha_i)_i$, and $p \in \Sigma$:
\begin{eqnarray}
\tau &\in& \eta(p) \\
\delta &\notin& \eta(p) \\
\alpha_i \in \eta(p)\ \forall\ i &\Leftrightarrow& \Pi_i\alpha_i \in \eta(p)
\end{eqnarray}
and for $p, q
\in \Sigma$ and $\alpha, \beta \in Q$ we have: \begin{eqnarray}
p < q &\Leftrightarrow& \eta(q) \subset \eta(p) \\ \alpha < \beta &\Leftrightarrow& \forall\ r \in \Sigma: \alpha \in \eta(r) {\rm
\ then}\ \beta \in \eta(r)
\end{eqnarray}
\end{proposition}
Proof: An infimum for the set $(\alpha_i)_i$ is a product test $\Pi_i\alpha_i$ as we have shown in proposition~\ref{compl}. It is also easy to see
that a product test
$\Pi_{\{\alpha_i <
\beta\
\forall i\}}\beta$ is a supremum for the family $(\alpha_i)_i$.\qed

\bigskip
\noindent
In general there is no a priori correspondence between properties and tests. Some properties can be tested and some tests give rise to properties. We have discussed this general situation in detail in (Aerts 1998) and will not repeat it here. In fact here, as this was also the case in the founding papers, we are interested in the situation where we consider only testable properties. And we will, as it was done in the founding papers, define properties as the equivalence classes of tests.

\begin{definition}
Consider a state test entity $S(\Sigma,Q,\eta)$. Two tests $\alpha,\beta \in Q$ are said to be `equivalent', $\alpha \approx \beta$, if both $\alpha < \beta$ and
$\beta < \alpha$ hold. In other words $\alpha \approx \beta$ iff for $p \in \Sigma$,
$\alpha
\in
\eta(p)
\Leftrightarrow \beta\in \eta(p)$.
\end{definition}If
$\alpha$ and $\beta$
are equivalent, they are considered to test the same property. That is why we will identify the properties of the entity with the equivalence classes of tests. 

\begin{definition} [property]
Consider a state test entity $S(\Sigma, Q, \eta)$. Let $\alpha \in Q$ be a test. The
`property' $a(\alpha)$ tested by $\alpha$ is defined to be the equivalence class of
$\alpha$ in $Q/_{\approx}$. In other words, \begin{equation}
a(\alpha)=\{\beta \in Q\ |\ \beta \approx \alpha\} \end{equation}
The set of all properties of the entity will be denoted ${\cal L}$, i.e.\ ${\cal
L}=Q/_{\approx}$.
\end{definition}
For the description of an entity by means of its states and properties we propose
state property systems, which were first defined in (Aerts 1998). We show that a unital product entity gives rise to a state property system. 

\begin{definition}\label{def:statprop}
We say that $(\Sigma,<,{\cal L},<,\wedge,\vee,\xi)$, or shorter $(\Sigma,{\cal L},\xi)$, is a
`state property system' if $(\Sigma,<)$ is a pre-ordered set,
$({\cal L},<, \wedge, \vee)$ is a complete lattice and $\xi$ is a function:
\begin{equation}
\xi: \Sigma \rightarrow {\cal P}({\cal L}) \end{equation}
such that for $p \in \Sigma$, $I$ the maximal element, $0$ the minimal element
of ${\cal L}$ and $(a_i)_i \in {\cal L}$, we have:
\begin{eqnarray}
I &\in& \xi(p) \\
0 &\not\in& \xi(p) \\
a_i \in \xi(p)\ \forall i &\Leftrightarrow& \wedge_ia_i \in \xi(p) \label{eq:xi_inf}
\end{eqnarray}
and for $p, q
\in \Sigma$ and $a, b\in {\cal L}$ we have: \begin{eqnarray}
p < q &\Leftrightarrow& \xi(q) \subset \xi(p) \label{eq:xi1}\\ a < b &\Leftrightarrow& \forall\ r \in \Sigma: a \in \xi(r) {\rm \ then}\ b \in \xi(r)
\label{eq:xi2}
\end{eqnarray}
\end{definition}
We remark that the reverse arrow of (\ref{eq:xi_inf}) follows from (\ref{eq:xi2}) and hence could be left out of the definition. Indeed, we clearly have
$\wedge_ia_i < a_j\
\forall j$ which means that $ \forall p \in \Sigma : \wedge_ia_i \in \xi(p)\Rightarrow a_j \in
\xi(p)\
\forall j$.

\begin{theorem}
Consider a unital product entity $S(\Sigma, Q, \eta)$. The triple $(\Sigma,{\cal
L},\xi)$ where
\begin{equation}
{\cal L} = \{ a(\alpha)\ \vert\ \alpha \in Q\} \end{equation}
is partially ordered by
\begin{equation}
a(\alpha) < a(\beta) \Leftrightarrow \alpha < \beta \ \ (\alpha,\beta \in Q) \end{equation}
and $\xi$ is the following function:
\begin{eqnarray}
\xi: \Sigma &\rightarrow& {\cal P}({\cal L}) \\ p &\mapsto& \xi(p) = \{a(\alpha)\ \vert\ \alpha \in \eta(p)\} \end{eqnarray}
is a state property system. The top and bottom element of $\cal L$ are given by
\begin{equation}
I = a(\tau)
\end{equation}
\begin{equation}
0 = a(\delta)
\end{equation}
where $\tau$ is a unit test and $\delta$ is a zero test. \end{theorem}
Proof: Let us prove that ${\cal L}$ is a complete lattice. The relation $<$ on ${\cal L}$ is well defined: $\alpha' \approx \alpha < \beta \approx \beta' \Rightarrow \alpha'<\beta'$. We clearly have that (${\cal L},<$) is a
pre-ordered set. We prove that $<$ is also antisymmetric. Consider two properties
$b,c \in {\cal L}$ such that $b<c$ and $c<b$. Then there exists $\epsilon,\gamma \in
Q$ such that $b=a(\epsilon)$ and $c=a(\gamma)$. $a(\epsilon) < a(\gamma)$ implies that
$\epsilon < \gamma$ and $a(\gamma) < a(\epsilon)$ implies that $\gamma < \epsilon$. This means that $\epsilon$ and $\gamma$ are equivalent. Consequently $a(\epsilon) = a(\gamma)$. This shows that (${\cal L},<$) is a partially ordered set.

Consider now an arbitrary set $(a_i)_i \in {\cal L}$. Then there exists a set
$(\alpha_i)_i \in Q$ such that $a_i=a(\alpha_i), \forall i$. Consider a product test
$\Pi_i\alpha_i$. Then $a(\Pi_i\alpha_i)$ is the infimum of the set $(a_i)_i$. Indeed, consider a state $p \in \Sigma$ such that $a(\Pi_i\alpha_i) \in \xi(p)$. Consequently $\alpha_i \in \eta(p), \forall i$. So we have $a(\alpha_i) \in \xi(p), \forall i$. This
shows that $a(\Pi_i\alpha_i) < a(\alpha_j), \forall j$. Suppose now that $a(\gamma) < a(\alpha_i), \forall i$ with $\gamma \in Q$ and consider a state $p \in \Sigma$ such that $a(\gamma) \in \xi(p)$. Then we have $a(\alpha_i) \in \xi(p) , \forall i$. Consequently we have that $\alpha_i \in \eta(p), \forall i$. This implies that $\Pi_i\alpha_i \in \eta(p)$ and so we have $a(\Pi_i\alpha_i) \in \xi(p)$. This shows that $a(\beta) < a(\Pi_i\alpha_i)$. Therefore $\cal L$ has arbitrary infima. It follows (and this is a result due to Birkhoff), that $\cal L$
has arbitrary suprema: for $(a_i)_i \in {\cal L}: \vee_ia_i=\bigwedge\{b \in {\cal
L}\ |\ a_i < b \forall i\}$. So $(\cal L,<)$ is a complete lattice.

For a unit test $\tau$ and a state $p$ we have that $\tau \in \eta(p)$. Consequently $I=a(\tau) \in \xi(p)$. For a zero test $\delta$ and a state $p$ we have that $\delta \notin \eta(p)$. This implies that $0=a(\delta) \notin \xi(p)$. 

Next we verify (\ref{eq:xi_inf}). Consider $(a_i)_i \in \cal L$ and a state $p$ such that $a_i \in \xi(p), \forall i$. Since there exists a set $(\alpha_i)_i \in Q $ such that $a_i=a(\alpha_i) \forall i$, we have that $\alpha_i \in \eta(p), \forall i$. Consequently $\Pi_i\alpha_i \in \eta(p)$. This implies that $a(\Pi_i\alpha_i) \in \xi(p)$. So we have that $\wedge_ia_i = \wedge_ia(\alpha_i) = a(\Pi_i\alpha_i) \in
\xi(p)$.

Equations (\ref{eq:xi1}) and (\ref{eq:xi2}) are easily verified. \qed

\section{State property systems and closure spaces} 

In this section we will investigate the state property systems and show that they are in natural correspondence with closure spaces. 

\begin{proposition}\label{prop:kappa}
Suppose that $(\Sigma,{\cal L},\xi)$ is a state property system. We introduce the
`Cartan map' $\kappa$:
\begin{equation}
\kappa: {\cal L} \rightarrow {\cal P}(\Sigma):a \mapsto \kappa(a) = \{p\in
\Sigma \ \vert\ a \in
\xi(p)\}
\end{equation}
For $a, b, (a_i)_i \in {\cal L}$ we have: \begin{eqnarray}
\kappa(I) &=& \Sigma \\
\kappa(0) &=& \emptyset \\
a < b &\Leftrightarrow& \kappa(a) \subset \kappa(b) \label{eq:kappa1}\\
\kappa(\wedge_ia_i) &=& \cap_i\kappa(a_i) \end{eqnarray}
It follows that $\kappa:{\cal L} \rightarrow (\kappa({\cal L}),\subset,\cap)$ is an isomorphism of
complete lattices.
\end{proposition}
Proof: Since $I \in \xi(p)\ \forall\ p \in \Sigma$, we have $\kappa(I) = \Sigma$. Since $0 \not\in \xi(p) \ \forall\ p \in \Sigma$ we have $\kappa(0) = \emptyset$. To prove (\ref{eq:kappa1}) just remark that (\ref{eq:xi2}) can be rewritten as
\begin{equation}
a < b \Leftrightarrow \forall\ r\in \Sigma : r\in \kappa(a) {\rm \ then}\ r\in \kappa(b)
\end{equation}
>From $\wedge_ia_i < a_j\
\forall j$ it follows that $\kappa(\wedge_ia_i) \subset \kappa(a_j)\ \forall j$. This yields $\kappa(\wedge_ia_i) \subset \cap_i\kappa(a_i)$. To prove the other inclusion, take $p \in \cap_i\kappa(a_i)$, then $p
\in \kappa(a_j)\ \forall j$. Hence $a_j \in \xi(p)\ \forall j$ which implies, by (\ref{eq:xi_inf}), that $\wedge_ia_i
\in \xi(p)$. From
this follows that $p \in \kappa(\wedge_ia_i)$. As a consequence we have $\cap_i\kappa(a_i) \subset \kappa(\wedge_ia_i)$.\qed

\medbreak
\noindent
To avoid misunderstandings we recall the definition of a closure space.
\begin{definition}\label{def:clos}
A `closure space' $(Z,{\cal G})$ consists of a set $Z$ and a family of subsets
${\cal G} \subset {\cal P}(Z)$
satisfying the following conditions:
\begin{eqnarray}
Z \in {\cal G} &,& \emptyset \in {\cal G} \\ (G_i)_i \in {\cal G} &\Rightarrow& \cap_iG_i \in {\cal G} \end{eqnarray}
If these conditions hold, we call ${\cal G}$ a `closure system' on $Z$. The `closure operator' corresponding to this closure space is defined as
\begin{equation}
cl: {\cal P}(Z) \rightarrow {\cal P}(Z): Y \mapsto \bigcap \{G \in {\cal G}\
\vert\ Y \subset G\}
\end{equation}
\end{definition}

\begin{theorem} \label{theor:sp-cls}
Suppose that $(\Sigma,{\cal L},\xi)$ is a state property system. Let us introduce
\begin{equation}
{\cal F} = \kappa({\cal L}) = \{\kappa(a)\ \vert\ a \in {\cal L}\} \end{equation}
Then ${\cal F}$ is a closure system on $\Sigma$. \end{theorem}
Proof: From the foregoing proposition follows that $\Sigma \in {\cal F}$ and
$\emptyset \in {\cal F}$. Consider $(F_i)_i \in {\cal F}$. Then there exists $(a_i)_i \in {\cal L}$ such that $\kappa(a_i) = F_i \ \forall i$.
We have $\kappa(\wedge_ia_i) = \cap_i\kappa(a_i) = \cap_iF_i$. This shows that $\cap_iF_i \in {\cal F}$.\qed 

\bigskip
\noindent
This theorem shows that to a state property system naturally corresponds a closure system on the set
of states, where the properties are represented by the closed subsets. We can also associate a state property system with any closure space. 

\begin{theorem} \label{theor:cls-sp}
Consider a closure space $(\Sigma,{\cal F})$. We introduce the following definitions for $F,G,
(F_i)_i \in {\cal F}$ and $p,q \in \Sigma$: \begin{eqnarray}
F < G &\Leftrightarrow& \ F \subset G \\ \wedge_iF_i &=& \cap_iF_i \\
\vee_iF_i &=& cl(\cup_iF_i)\\
\xi: \Sigma \rightarrow {\cal P}({\cal F})&:& p \mapsto \{F \in {\cal F}\
\vert\ p \in F\}\\
p < q &\Leftrightarrow& \xi(q) \subset \xi(p) \label{eq:clos_preo} \end{eqnarray}
Then $(\Sigma, <, {\cal F}, <, \wedge, \vee, \xi)$ is a state property system.
\end{theorem}
Proof: It is easy to show that $({\cal F}, <, \wedge, \vee)$ is a complete lattice, with maximal element $I=\Sigma$ and minimal element $0 =
\emptyset$. It is trivial
to verify that (\ref{eq:clos_preo}) defines a pre-order on $\Sigma$. Clearly, we have $I
\in \xi(p), 0 \not\in \xi(p)\ \forall\ p \in \Sigma$. Next, suppose that
$F_i
\in \xi(p)\ \forall i$. This means that $p \in F_i\ \forall i$ or $p \in \cap_iF_i$. As a consequence we have $\wedge_iF_i=\cap_iF_i \in \xi(p)$. Finally we verify (\ref{eq:xi2}). Let
$F, G
\in {\cal L}$. We then have $F < G \Leftrightarrow F\subset G \Leftrightarrow (p\in F \Rightarrow
p\in G) \Leftrightarrow (F \in \xi(p) \Rightarrow G \in \xi(p))$ and we are done. \qed

\section{The morphisms}

Theorem~\ref{theor:sp-cls} and theorem~\ref{theor:cls-sp} show that there is a
straightforward correspondence between state property systems and closure spaces. We can extend this correspondence to ``natural'' morphisms of these
two structures. In this section we introduce morphisms of state property systems and show their
connection to continuous maps between closure spaces.

Consider two state property systems
$(\Sigma, {\cal L},\xi)$ and
$(\Sigma',{\cal L}',\xi')$. As we have explained in section~\ref{sec:phys}, these
state property systems respectively describe entities $S$ and $S'$. We will arrive at the notion of morphism by analyzing the situation where the entity $S$ is a sub-entity of the entity $S'$. In that case, the following three natural requirements should be satisfied:

\smallskip
\noindent i) If the entity $S'$ is in a state $p'$ then the state $m(p')$ of
$S$
is determined. This defines a function $m$ from the set of states of $S'$ to the set of states of $S$; 

\smallskip
\noindent ii) If we consider a property $a$ of the entity $S$, then to
$a$
corresponds a property $n(a)$ of the ``bigger'' entity $S'$. This defines a function $n$ from
the set of properties of $S$ to the set of properties of $S'$; 

\smallskip
\noindent
iii) We want $a$ and $n(a)$ to be two descriptions of the ``same'' property of $S$, once
considered as an entity on itself, once as a sub-entity of $S'$. In other words we want $a$ and
$n(a)$ to be actual at once. This means that for a state
$p'$ of $S'$ (and a
corresponding state $m(p')$ of $S$) we want the following ``covariance principle'' to hold:
\begin{equation}
a \in \xi(m(p')) \Leftrightarrow n(a) \in \xi'(p') \end{equation}
We are now ready to present a formal definition of a morphism of state property systems.

\begin{definition} \label{def:morphism}
Consider two state property systems $(\Sigma, {\cal L}, \xi)$ and $(\Sigma',{\cal L}', \xi')$. We
say that
\begin{equation}
(m,n):(\Sigma',{\cal L}',\xi') \longrightarrow (\Sigma,{\cal L},\xi) \end{equation}
is a `morphism' (of state property systems) if $m$ is a function:
\begin{equation}
m: \Sigma' \rightarrow \Sigma
\end{equation}
and $n$ is a function:
\begin{equation}
n: {\cal L} \rightarrow {\cal L}'
\end{equation}
such that for $a \in {\cal L}$ and $p' \in \Sigma'$ the following holds:
\begin{equation} \label{eq:covar1}
a \in \xi(m(p')) \Leftrightarrow n(a) \in \xi'(p') \end{equation}
\end{definition}
The following is an elegant rewriting of this definition: \begin{proposition}
Consider two state property systems $(\Sigma, {\cal L},\xi)$ and $(\Sigma',{\cal L}',\xi')$. Two
functions $m:\Sigma' \rightarrow \Sigma$ and $n:{\cal L} \rightarrow {\cal L}'$ define a morphism $(m,n): (\Sigma',{\cal L}',\xi') \rightarrow (\Sigma,{\cal L},\xi)$ if and only if we have
\begin{equation}
\xi \circ m = n^{-1} \circ \xi' \label{eq:covar2} \end{equation}
where $n^{-1}:{\cal P}({\cal L}') \rightarrow {\cal P}({\cal L}): F' \mapsto n^{-1}(F')=\{a \in {\cal
L} \ \vert\ n(a) \in F'\}$.
\end{proposition}
The next proposition gives some properties of morphisms. \begin{proposition}
Consider two state property systems $(\Sigma,{\cal L},\xi)$ and $(\Sigma',{\cal L}',\xi')$
connected by a morphism $(m,n): (\Sigma',{\cal L}',\xi') \rightarrow (\Sigma,{\cal L},\xi)$. For
$p', q' \in \Sigma'$ and $a, b,(a_i)_i \in {\cal L}$ we have: \begin{eqnarray}
p' < q' &\Rightarrow& m(p') < m(q') \\
a < b &\Rightarrow& n(a) < n(b) \\
n(\wedge_ia_i) &=& \wedge_in(a_i) \label{eq:ninf} \\ n(I) &=& I' \\
n(0) &=& 0'
\end{eqnarray}
\end{proposition}
Proof:
Suppose that $p' < q'$. We then have $\xi'(q') \subset \xi'(p')$. From this follows
that $n^{-1}(\xi'(q'))
\subset n^{-1}(\xi'(p'))$. Through (\ref{eq:covar2}) this yields $\xi(m(q'))
\subset
\xi(m(p'))$, whence $m(p') < m(q')$.

Next consider $a < b$ and let $r' \in \Sigma'$ be such that $n(a) \in \xi'(r')$. Then we have
$a \in \xi(m(r'))$ and, since $a < b$, this yields $b \in \xi(m(r'))$. From this follows that $n(b) \in \xi'(r')$. So we have shown that $n(a) < n(b)$.

>From $\wedge_ia_i < a_j\ \forall j$ we obtain 
$n(\wedge_ia_i) < n(a_j)\ \forall j$. This yields $n(\wedge_ia_i) < \wedge_in(a_i)$. We still
have to show that
$\wedge_in(a_i) < n(\wedge_ia_i)$. Let $r' \in \Sigma'$ be such that $\wedge_in(a_i) \in \xi'(r')$.
This implies that
$n(a_j) \in \xi'(r')\ \forall j$ (use (\ref{eq:xi2})). But from this we obtain $a_j \in \xi(m(r'))\ \forall j$ and hence $\wedge_ia_i \in \xi(m(r'))$. As a consequence we have
$n(\wedge_ia_i) \in \xi'(r')$. But then we have shown that $\wedge_in(a_i) < n(\wedge_ia_i)$.

We clearly have $n(I) < I'$. For all $r'\in \Sigma'$, we have $I \in \xi(m(r'))$ and hence $n(I) \in
\xi'(r')$. Through (\ref{eq:xi2}) this implies $I' < n(I)$, whence $n(I)=I'$.
Trivially $0' < n(0)$. Suppose $n(0) < 0'$ does not hold. Then the contraposition of (\ref{eq:xi2})
says there is an $r'\in\Sigma'$ such that $n(0) \in \xi'(r')$. This would imply $0 \in \xi(m(r'))$ which is impossible. Therefore we have proven $n(0)=0'$. \qed

\begin{proposition}\label{prop:kappanat} Suppose that we have a morphism of state property systems $(m,n):(\Sigma',{\cal L}',\xi') \rightarrow (\Sigma,{\cal L},\xi)$. Consider the Cartan maps $\kappa$ and $\kappa'$ that connect these state property systems to their corresponding closure spaces $(\Sigma, {\cal F})$ and $(\Sigma', {\cal F}')$, as was done in theorem~\ref{theor:sp-cls}. For $a \in {\cal
L}$ we have:
\begin{equation}
m^{-1}(\kappa(a)) = \kappa'(n(a))
\end{equation}
\end{proposition}
Proof: We have: $p' \in m^{-1}(\kappa(a)) \Leftrightarrow m(p') \in \kappa(a) \Leftrightarrow a \in \xi(m(p')) \Leftrightarrow n(a) \in \xi'(p') \Leftrightarrow p' \in \kappa'(n(a))$.\qed 

\bigskip
\noindent We can now connect morphisms of state property systems to continuous maps (morphisms of closure
spaces).
\begin{proposition} \label{theor:mor-cont} Suppose that we have a morphism of state property systems $(m,n):(\Sigma',{\cal L}',\xi') \rightarrow (\Sigma,{\cal L},\xi)$. If $(\Sigma,{\cal F})$ and
$(\Sigma', {\cal F}')$
are the closure spaces corresponding to these state property systems (cf.\ theorem~\ref{theor:sp-cls}),
then $m:(\Sigma',{\cal F}')\rightarrow (\Sigma,{\cal F})$ is continuous.
\end{proposition}
Proof: Take a closed subset $F \in {\cal F}$. Then there is an $a \in {\cal L}$ such that $\kappa(a) = F$. From the foregoing proposition we have $m^{-1}(F)
= m^{-1}(\kappa(a)) = \kappa'(n(a)) \in {\cal F}'$. This proves our claim.\qed

\begin{proposition}\label{theor:cont-mor} Suppose we have two closure spaces $(\Sigma, {\cal F})$ and $(\Sigma', {\cal F}')$ and a continuous map $m: \Sigma' \rightarrow \Sigma$. Consider the state property systems $(\Sigma,{\cal F},\xi)$ and
$(\Sigma',{\cal F}',\xi')$ corresponding to these two closure systems, as proposed in theorem~\ref{theor:cls-sp}. Then $(m,m^{-1})$ is a morphism from $(\Sigma',{\cal F}',\xi')$ to
$(\Sigma,{\cal F},\xi)$.
\end{proposition}
Proof: Continuity yields that $m^{-1}$ is a function from ${\cal F}$ to ${\cal F}'$. Let us now show formula (\ref{eq:covar1}) using the
definition of $\xi'$ and $\xi$ as put forward in theorem~\ref{theor:cls-sp}. For $F \in {\cal F}$ and $p'\in \Sigma'$ we have
$F \in \xi(m(p'))
\Leftrightarrow m(p')
\in F
\Leftrightarrow p
\in m^{-1}(F)
\Leftrightarrow m^{-1}(F) \in \xi'(p')$. \qed 

\section {An equivalence of categories}
The previous section demonstrates that there is a strong connection between state property systems with their morphisms and closure spaces with continuous maps. In this section we formalize this connection into an equivalence of categories. We suppose the reader to be familiar
with basic category theory and refer him or her who isn't to (Borceux 1994). 

We will introduce the categories, but before doing so, we define the composition of morphisms of
state property systems.
\begin{definition}
Given two morphisms of state property systems $(m_1,n_1):(\Sigma_1,{\cal L}_1,\xi_1) \rightarrow (\Sigma_2,{\cal L}_2,\xi_2)$ and $(m_2,n_2):(\Sigma_2,{\cal L}_2,\xi_2)\rightarrow(\Sigma_3,{\cal L}_3,\xi_3)$ their composite is defined as
\begin{equation}
(m_2,n_2)\circ(m_1,n_1) = (m_2 \circ m_1, n_1\circ n_2) \end{equation}
\end{definition}
\begin{proposition}
Given two morphisms of state property systems $(m_1,n_1):(\Sigma_1,{\cal L}_1,\xi_1) \rightarrow (\Sigma_2,{\cal L}_2,\xi_2)$ and $(m_2,n_2):(\Sigma_2,{\cal L}_2,\xi_2)\rightarrow(\Sigma_3,{\cal
L}_3,\xi_3)$ their composite $(m_2,n_2)\circ(m_1,n_1): (\Sigma_1,{\cal L}_1,\xi_1) \rightarrow (\Sigma_3,{\cal
L}_3,\xi_3)$ is again a morphism of state property systems. \end{proposition}
Proof: We prove our claim by checking formula (\ref{eq:covar2}). We have
$\xi_3 \circ (m_2 \circ
m_1) = (\xi_3 \circ m_2) \circ m_1 = (n_2^{-1} \circ \xi_2) \circ m_1 = n_2^{-1}\circ (\xi_2 \circ
m_1) = n_2^{-1} \circ (n_1^{-1} \circ \xi_1) = (n_1 \circ n_2)^{-1} \circ
\xi_1$, which proves
the assertion. \qed

\bigskip \noindent The following does not need a proof. \begin{proposition}
The composition of morphisms of state property systems is associative and given a morphism
$(m,n):(\Sigma',{\cal L}',\xi') \rightarrow (\Sigma,{\cal L},\xi)$ the following equalities hold:
\begin{eqnarray}
(m,n)\circ (id_{_{\Sigma'}},id_{_{\cal L'}}) &=& (m,n)\\ (id_{_{\Sigma}},id_{_{{\cal L}}})\circ (m,n) &=& (m,n) \end{eqnarray}
\end{proposition}
Having these results under our belt, we can safely state the following definitions.
\begin{definition}
We call ${\bf SP}$ the category of state property systems (definition~\ref{def:statprop}) with
their morphisms (definition~\ref{def:morphism}) and ${\bf Cls}$ is the category of closure spaces (definition~\ref{def:clos}) with continuous maps. \end{definition}
Let us introduce the functors which will establish the equivalence of categories.
\begin{theorem}
\label{corrF}
The correspondence $F : {\bf SP} \longrightarrow {\bf Cls}$ consisting of
\newline (1) the mapping
\begin{eqnarray}
|{\bf SP}| &\rightarrow& |{\bf Cls}| \\
(\Sigma,{\cal L},\xi) &\mapsto& F(\Sigma,{\cal L},\xi) \end{eqnarray}
where $F(\Sigma,{\cal L},\xi)$ is the closure space $(\Sigma,{\cal F})$ given by
theorem~\ref{theor:sp-cls};
\newline (2) for every pair of objects $(\Sigma,{\cal L},\xi),(\Sigma',{\cal L}',\xi')$
of $\bf SP$ the mapping
\begin{eqnarray}
{\bf SP}((\Sigma',{\cal L}',\xi'),(\Sigma,{\cal L},\xi)) &\rightarrow& {\bf
Cls}(F(\Sigma',{\cal L}',\xi'),F(\Sigma,{\cal L},\xi)) \\ (m,n) &\mapsto& m
\end{eqnarray}
is a covariant functor.
\end{theorem}
Proof: This is, apart from some minor checks, a consequence of theorem~\ref{theor:sp-cls} and
proposition~\ref{theor:mor-cont}. \qed

\begin{theorem} \label{corrG}
The correspondence $G : {\bf Cls} \longrightarrow {\bf SP}$ consisting of
\newline (1) the mapping
\begin{eqnarray}
|{\bf Cls}| &\rightarrow& |{\bf SP}| \\
(\Sigma,{\cal F}) &\mapsto& G(\Sigma,{\cal F}) \end{eqnarray}
where $G(\Sigma,{\cal F})$ is the state property system $(\Sigma, {\cal F},\xi)$ given by
theorem~\ref{theor:cls-sp};
\newline (2) for every pair of objects $(\Sigma,{\cal F}),(\Sigma', {\cal F}')$
of $\bf Cls$ the mapping
\begin{eqnarray}
{\bf Cls}((\Sigma',{\cal F}'),(\Sigma,{\cal F})) &\rightarrow& {\bf SP}(G(\Sigma',{\cal F}'),G(\Sigma,{\cal F})) \\ m &\mapsto& (m,m^{-1})
\end{eqnarray}
is a covariant functor.
\end{theorem}
Proof: This is, apart from some minor checks, a consequence of theorem~\ref{theor:cls-sp} and
proposition~\ref{theor:cont-mor}. \qed

\bigskip\noindent
Next we characterize the isomorphisms of $\bf SP$. \begin{proposition}\label{prop:statpropiso} A morphism $(m,n) \in {\bf SP}((\Sigma',{\cal L}',\xi'),(\Sigma,{\cal L},\xi))$ is an isomorphism if
and only if $m:\Sigma' \rightarrow \Sigma$ and $n: {\cal L}\rightarrow {\cal L}'$ are bijective.
\end{proposition}
Proof: Let $(m,n)$ be an isomorphism. The fact that it has a right inverse implies that $m$ is
surjective and that $n$ is injective. On the other hand we conclude from the existence of a left
inverse that $m$ is injective and that $n$ is surjective. 

Now, let $(m,n)$ be a morphism with $m$ and $n$ bijective. Let $m^{-1}:\Sigma \rightarrow \Sigma'$
and $n^{-1}:{\cal L}' \rightarrow {\cal L}$ be the inverses of $m$ and $n$. We show that
$(m^{-1},n^{-1})$ is a morphism, using (\ref{eq:covar2}). From $\xi \circ m= n^{-1} \circ \xi'$ we
obtain $\xi' = n \circ \xi \circ m$ , where $n^{-1}:{\cal P}({\cal L}') \rightarrow {\cal P}({\cal L})$ and $n:{\cal P}({\cal L}) \rightarrow {\cal P}({\cal L}'): T \mapsto n(T) = \{n(a) \ \vert\
a \in T \}$. This implies $\xi' \circ m^{-1} = n \circ \xi \circ m \circ m^{-1} = n\circ \xi$,
which proves our assertion. \qed

\bigskip
\noindent We have arrived at
\begin{theorem}[Equivalence of $\bf SP$ and $\bf Cls$] \label{theor:equiv} The functors $F:{\bf SP}\rightarrow {\bf Cls}$ and $G:{\bf Cls}
\rightarrow {\bf SP}$ establish an
equivalence of categories. Moreover $F\circ G = Id_{_{\bf Cls}}$. \end{theorem}
Proof: Step 1: $G \circ F$. Given an object $(\Sigma,{\cal L},\xi) \in |{\bf SP}|$, we have
$GF(\Sigma,{\cal L},\xi) = (\Sigma, \kappa({\cal L}),\overline{\xi})$ where $\kappa:{\cal
L}\rightarrow{\cal P}(\Sigma)$ is the Cartan map defined in proposition~\ref{prop:kappa} and
\begin{eqnarray}
\overline{\xi}: \Sigma &\rightarrow& {\cal P}(\kappa({\cal L})) \\ 
p &\mapsto& \{\kappa(a)\ \vert\ a\in {\cal L}, p\in \kappa(a)\}= \{\kappa(a)\ \vert\ a \in \xi(p)\}
\end{eqnarray}
Given a morphism $(m,n) \in {\bf SP}((\Sigma',{\cal L}',\xi'),(\Sigma, {\cal L},\xi))$, we obtain
$GF(m,n)=(m,m^{-1}) \in {\bf SP}((\Sigma', \kappa'({\cal L}'),\overline{\xi'}),(\Sigma, \kappa({\cal L}),\overline{\xi})$.

Step 2: $GF \cong Id_{_{\bf SP}}$. For any object $(\Sigma,{\cal L}, \xi)
\in |{\bf SP}|$, define
\begin{eqnarray}
\varepsilon_{_{(\Sigma,{\cal L},\xi)}}:GF(\Sigma,{\cal L},\xi) &\rightarrow& (\Sigma,{\cal
L},\xi)\\
\varepsilon_{_{(\Sigma,{\cal L},\xi)}} &=& (id_{_{\Sigma}},\kappa) \end{eqnarray}
Then $\varepsilon = (\varepsilon_{_{(\Sigma,{\cal L},\xi)}}) : GF \rightarrow Id_{_{\bf SP}}$ is a natural isomorphism. First we verify that $\varepsilon_{_{(\Sigma,{\cal L},\xi)}}$ is a morphism of $\bf SP$. Indeed, for $a
\in {\cal L}, p \in
\Sigma$ we have $\kappa(a) \in \overline{\xi}(p) \Leftrightarrow p \in
\kappa(a) \Leftrightarrow a
\in \xi(p)=\xi(id_{_{\Sigma}}p)$. To show that $(id_{_{\Sigma}}, \kappa)$ is an isomorphism, we only
have to prove that $\kappa:{\cal L}\rightarrow\kappa({\cal L})$ is bijective
(proposition~\ref{prop:statpropiso}) and this follows from proposition~\ref{prop:kappa}. The
naturality of $\varepsilon$ is an immediate consequence of proposition~\ref{prop:kappanat}.

Step 3: $FG = Id_{_{\bf Cls}}$. For the morphisms this is trivial: \begin{equation}
m \stackrel{G}{\longmapsto} (m,m^{-1}) \stackrel{F}{\longmapsto} m \end{equation}
where m is a morphism of $\bf Cls$. Now
consider an arbitrary closure space $(\Sigma,{\cal F})$. Then $G(\Sigma,{\cal F})=(\Sigma,{\cal
F},\xi)$ where $\xi(p)=\{F \in {\cal F}\ \vert\ p\in F\}$. Hence the corresponding Cartan map is
given by
\begin{eqnarray}
\kappa:{\cal F} &\rightarrow& {\cal P}(\Sigma) \\ F &\mapsto& \{p \in \Sigma\ \vert\ F \in \xi(p)\} = \{p \in \Sigma\ \vert\ p \in F\} = F
\end{eqnarray}
This implies $FG(\Sigma,{\cal F})=F(\Sigma,{\cal F},\xi) = (\Sigma, \kappa({\cal F}))= (\Sigma,{\cal
F})$.\qed

\section{The first axiom: state determination and T$_{_0}$ separation}\label{sec:T0}

In the founding papers the state $p\in
\Sigma$ of an entity $S$ is identified with the set of all properties $a \in {\cal L}$ it makes actual. In this section, we investigate the consequences this assumption has on state property systems and closure spaces.

Let $S(\Sigma,Q,\eta)$ be a unital product entity and let
$(\Sigma,{\cal L},\xi)$ be its state property system. Remember that for a state $p\in \Sigma$ we have that
\begin{equation}
\xi(p)=\{a \in {\cal L}\ |\ a\mbox { is actual when }S\mbox{ is in state }p\}
\end{equation}
Hence the demand that a state $p$ be completely determined by the set of all properties it makes actual, i.e.\ by $\xi(p)$, is mathematically expressed by:
\smallbreak
\centerline{``$\xi:\Sigma \rightarrow {\cal P}({\cal L})$ is injective.''} \smallbreak
\begin{definition} A closure space $(Z,{\cal G})$ is called `T$_{_0}$' if for $x,y \in Z$ we have $cl(x)=cl(y) \Rightarrow x=y$, where $cl(x)$ is the usual notation for $cl(\{x\})$. \end{definition}
Let us give some equivalent conditions to the injectivity of $\xi$.

\begin{proposition} \label{prop:xi_inj}
Let $S(\Sigma,Q,\eta)$ be a unital product entity and let $(\Sigma,{\cal L},\xi)$ be the state property system it generates. The following are equivalent:
\par (1) $\xi:\Sigma \rightarrow {\cal P}({\cal L})$ is injective; \par (2) the pre-order $<$ on $\Sigma$ is a partial order; \par (3) $\eta:\Sigma \rightarrow{\cal P}(Q)$ is injective; \par (4) $(\Sigma,{\cal F}) = F(\Sigma,{\cal L},\xi)$ is a T$_{_0}$ closure space.
\end{proposition}
Proof: (1$\Leftrightarrow$2) Remember
that we have for $p,q \in \Sigma$: $p
< q$ iff
$\xi(q)
\subset
\xi(p)$. Hence $p < q$ and $q < p$ is equivalent to $\xi(q) \subset \xi(p)$ and $ \xi(p) \subset \xi(q)$ (or $\xi(p)=\xi(q)$). It follows that the injectivity of $\xi$ (i.e.\ $\xi(p)=\xi(q)
\Rightarrow p=q$) is equivalent to the antisymmetry of $<$ (i.e.\ $p < q$ and $q < p
\Rightarrow p=q$).

(1$\Leftrightarrow$3) This is an immediate consequence of $\eta(q) \subset \eta (p)
\Leftrightarrow p < q \Leftrightarrow \xi(q) \subset \xi(p)$, where the first `$\Leftrightarrow$' is the definition of $<$ and where $p,q \in \Sigma$.

(1$\Rightarrow$4) Suppose $p,q \in \Sigma$ are such that $cl(p)=cl(q)$. From the definition of $(\Sigma,{\cal F})$ we have that $cl(p) = \bigcap_{p
\in \kappa(a)}\kappa (a)=\bigcap_{a \in \xi(p)}\kappa (a)$, where $\kappa: {\cal L}\rightarrow {\cal P}(\Sigma)$ is the Cartan map defined in proposition \ref{prop:kappa}. Hence we have $p \in \bigcap_{a \in \xi(p)}\kappa(a) =
\bigcap_{a \in \xi(q)}\kappa (a)\ni q$. This yields that $p \in \kappa(a)$ for every $a
\in \xi(q)$, or in other words, $a \in \xi(q) \Rightarrow a\in \xi(p)$. This shows
$\xi(q) \subset \xi(p)$. Similarly $q \in \bigcap_{a \in \xi(p)}\kappa (a)$ gives $\xi(p)
\subset \xi(q)$. So we have $\xi(p)=\xi(q)$, whence by (1) $p=q$ holds.

(4$\Rightarrow$1) Consider $p,q \in \Sigma$ with $\xi(p)=\xi(q)$. Then $cl(p)=\bigcap_{a\in \xi(p)}\kappa(a) = \bigcap_{a\in \xi(q)}\kappa(a) = cl(q)$. Since $(\Sigma,{\cal F})$ is T$_{_0}$ (4) we have $p=q$. \qed 

\bigskip
\noindent
The following terminology is taken from (Aerts 1994). \begin{definition}{\bf (state determined entity)} We call a state test entity $S(\Sigma,Q,\eta)$ `state determined' if $\eta:\Sigma
\rightarrow {\cal P}(Q)$ is injective. We will call a unital product entity $S(\Sigma,Q,\eta)$ a `state determined entity' if it is state determined. A state property system $(\Sigma,{\cal L},\xi)$ is a `state determined state property system' if $\xi$ is injective. \end{definition}

\begin{definition}
We define $\bf SP_{_0}$ as the subcategory of $\bf SP$ where the objects are given by
\begin{equation}
|{\bf SP_{_0}}| = \{(\Sigma,{\cal L},\xi) \in |{\bf SP}| : \xi \mbox{ is injective}\}
\end{equation}
and the morphisms by
\begin{equation}
{\bf SP_{_0}}((\Sigma',{\cal L}',\xi'),(\Sigma,{\cal L},\xi)) = {\bf SP}((\Sigma',{\cal L}',\xi'),(\Sigma,{\cal L},\xi)) \label{eq:morph0} \end{equation}
where $(\Sigma,{\cal L},\xi),(\Sigma',{\cal L}',\xi') \in {\bf SP_{_0}}$. So $\bf SP_{_0}$ is the category of state determined state property systems. Similarly we will use $\bf Cls_{_0}$ for the category of T$_{_0}$ closure spaces with continuous maps as morphisms. \end{definition}
Clearly $\bf Cls_{_0}$ is an isomorphism-closed subcategory of $\bf Cls$. We prove that the same holds for $\bf SP_{_0}$. 

\begin{proposition}
The category $\bf SP_{_0}$ is an isomorphism-closed subcategory of $\bf SP$: if $(\Sigma',{\cal L}',\xi')
\in {\bf SP_{_0}}$, $(\Sigma,{\cal L},\xi) \in {\bf SP}$ and $(m,n):(\Sigma',{\cal L}',\xi') \rightarrow (\Sigma,{\cal L},\xi) $ is an isomorphism of $\bf SP$, then
$(m,n)$ is an isomorphism of $\bf SP_{_0}$, in particular $(\Sigma, {\cal L},\xi) \in {\bf SP_{_0}}$. \end{proposition}
Proof: By equation (\ref{eq:morph0}) we only have to show that $(\Sigma,{\cal L},\xi)
\in {\bf SP_{_0}}$, i.e.\ that $\xi$ is injective. Suppose $\xi(p)=\xi(q)$ holds for some $p,q \in \Sigma$. Put $p'=m^{-1}(p), q'=m^{-1}(q)$. We show
$\xi'(p')=\xi'(q')$. Indeed, $a = n(n^{-1}(a)) \in \xi'(p') \Leftrightarrow n^{-1}(a) \in \xi(m(p'))=\xi(p)=\xi(q)=\xi(m(q')) \Leftrightarrow a= n(n^{-1}(a)) \in \xi'(q')$. Since $\xi$ is injective, this implies $p'=q'$, whence
$p=q$. \qed

\bigskip
\noindent
We also have the following:
\begin{proposition} \label{prop:G0}
Let $(\Sigma,{\cal F})$ be a closure space. Let $G: {\bf Cls} \rightarrow {\bf SP}$ be the functor defined in theorem~\ref{corrG}. Then
\begin{equation}
(\Sigma,{\cal F})
\in |{\bf Cls_{_0}}| \Longleftrightarrow G(\Sigma,{\cal F})=(\Sigma, \cal F,\xi) \in |{\bf SP_{_0}}|
\end{equation}
where, as in theorem~\ref{theor:cls-sp}, $\xi:\Sigma \rightarrow {\cal P}({\cal F}): p
\mapsto \{F \in {\cal F}\ |\ p \in F\}$. \end{proposition}
Proof: ($\Rightarrow$) Suppose $cl(p) = cl(q)$ implies $p=q$ for all $p,q \in \Sigma$. We have to show that $\xi$ is injective. Suppose $\xi(p) = \xi(q)$ for some $p, q\in
\Sigma$. Then $cl(p) = \bigcap \xi(p) = \bigcap \xi(q) = cl(q)$. This yields $p=q$.

\noindent ($\Leftarrow$) If $G(\Sigma,{\cal F}) = (\Sigma, \cal F,\xi) \in |{\bf SP_{_0}}|$,
then $\xi$ is injective. By theorem~\ref{theor:equiv} and proposition ~\ref{prop:xi_inj}
$(\Sigma,{\cal F}) = FG(\Sigma,{\cal F}) \in |{\bf Cls_{_0}}|$. \qed 

\bigskip
\noindent
We can now prove the following:
\begin{theorem}{\bf (equivalence of $\bf SP_{_0}$ and $\bf Cls_{_0}$)} \label{theor:equiv0} The covariant functors $F$ and $G$ (see theorem \ref{corrF} and theorem
\ref{corrG}) restrict and corestrict to functors $F:{\bf SP_{_0}}\rightarrow {\bf
Cls_{_0}}$ and $G: {\bf Cls_{_0}} \rightarrow {\bf SP_{_0}}$, which establish an
equivalence of categories.
\end{theorem}
Proof: This is an immediate consequence of theorem~\ref{theor:equiv} and propositions~\ref{prop:xi_inj} and \ref{prop:G0}. \qed

\section{States as strongest actual properties} \label{sec:embstate} 

Let $S(\Sigma,Q,\eta)$ be a state determined entity and let $(\Sigma,{\cal L},\xi)$ be its state determined state property system. Then it is possible to identify a state $p$ of $S$ with the strongest property it makes actual, i.e.\ with $\bigwedge\xi(p) \in {\cal L}$. As a consequence, one can embed
$(\Sigma,<)$ into $({\cal L},<,\wedge,\vee)$ as an order-generating subset. This engenders another equivalence of categories.

We start by embedding $(\Sigma,<)$ into $({\cal L},<)$. \begin{theorem} \label{prop:embstate}
Let $(\Sigma,{\cal L},\xi)$ be a state property system. The following are equivalent:
\par (1) $(\Sigma,{\cal L},\xi)$ is a state determined state property system; \par (2) If we define
\begin{equation}
s_{\xi}:\Sigma \rightarrow {\cal L}: p \mapsto \bigwedge\xi(p) \end{equation}
then $s_{\xi}$ is injective and for $p,q \in \Sigma$ we have \begin{equation}
p < q \Leftrightarrow s_{\xi}(p) < s_{\xi}(q) \label{eq:ixi} \end{equation}
Therefore, if $(\Sigma,{\cal L},\xi)$ is state determined, then $s_{\xi}$ is isotone and injective and
$(\Sigma,<)$ can be considered as a sub-poset of $({\cal L},<)$. We will use the notation $\Sigma^{\xi} = s_{\xi}(\Sigma)$. \end{theorem}
Proof: (2$\Rightarrow$1) Equation (\ref{eq:ixi}) and the injectivity of $s_{\xi}$ imply that $(\Sigma,<)$ is a poset, whence, through proposition~\ref{prop:xi_inj},
$\xi$ is injective.

(1$\Rightarrow$2) We first verify (\ref{eq:ixi}). Suppose $p,q \in \Sigma$. Then $p < q \Leftrightarrow \xi(q) \subset \xi(p) \Leftrightarrow [s_{\xi}(q),I] \subset [s_{\xi}(p),I] \Leftrightarrow s_{\xi}(p) < s_{\xi}(q)$. Since the injectivity of $\xi$ implies that $(\Sigma,<)$ is a poset, the injectivity of $s_{\xi}$ follows from (\ref{eq:ixi}). \qed

In the proof of theorem \ref{prop:Hob} we will use the following result. 

\begin{proposition}
Let $(\Sigma,{\cal L},\xi)$ be a state property system. If we use the notations of Theorem \ref{prop:embstate} for a state $p \in \Sigma$ and a property $a \in \cal L$, we have the following equivalence. \begin{equation}
a \in \xi(p) \Leftrightarrow s_{\xi}(p) < a \end{equation}
\end{proposition}
Proof: ($\Rightarrow$) This implication follows immediately from the definition of $s_{\xi}$. \newline
($\Leftarrow$) Suppose $s_{\xi}(p) < a$. Applying (\ref{eq:xi2}) for state $p$ we have that $s_{\xi}(p) = \bigwedge\xi(p) \in \xi(p)$ implies that $a \in \xi(p)$. \qed

\begin{theorem} \label{prop:Hob}
Let $(\Sigma,{\cal L},\xi)$ be a state property system. Then $0 \not\in \Sigma^{\xi}$ and $\Sigma^{\xi}$ is an order-generating subset of $\cal L$: for every $a \in {\cal L}$ we have \begin{equation}
a= \bigvee\{x \in \Sigma^{\xi}\ |\ x < a\} \label{eq:og} \end{equation}
\end{theorem}
Proof: Since $0 \not\in \xi(p)$ for every $p$, we have $0 \not\in \Sigma^{\xi}$. The `$>$' of equation (\ref{eq:og}) is trivial. To show `$<$', we will use equation (\ref{eq:xi2}). So, take $p\in \Sigma$ such that $a \in \xi(p)$. Then $s_{\xi}(p) < a$ and hence $s_{\xi}(p) \in \{x \in \Sigma^{\xi}\ |\ x < a\}$. This implies $s_{\xi}(p) < \bigvee\{x \in \Sigma^{\xi}\ |\ x < a\}$, or $\bigvee\{x \in \Sigma^{\xi}\ |\ x < a\} \in \xi(p)$. This proves $a<\bigvee\{x \in \Sigma^{\xi}\ |\ x < a\}$. \qed

\bigskip
\noindent
We introduce some notation, which should make clear what our intentions are.
\begin{definition} \label{def:Hob}
Let $(\Sigma,{\cal L},\xi) \in {\bf SP}$. Then we put \begin{equation}
H(\Sigma,{\cal L},\xi) := (\Sigma^{\xi},{\cal L}) \end{equation}
\end{definition}
Now, let us try to go ``back''. First we introduce some terminology.

\begin{definition}
We call $(\Sigma,{\cal L})$ a `based complete lattice' if $\cal L$ is a complete lattice and $\Sigma \subset {\cal L}$ is an order-generating subset not containing $0$. \end{definition}
>From theorem \ref{prop:Hob} it follows that for every state property system 
$(\Sigma,{\cal L},\xi)$,
$H(\Sigma,{\cal L},\xi)$ is a based complete lattice.

\begin{theorem} \label{prop:Kob}
Let $(\Sigma,{\cal L})$ be a based complete lattice. If we put for $p, q\in \Sigma$:
\begin{equation}
p < q \Leftrightarrow p \prec q \mbox{ ($\prec$ is the order of ${\cal L}$)}
\end{equation}
and
\begin{eqnarray}
\xi:\Sigma &\rightarrow& {\cal P}({\cal L}) \\ p &\mapsto& \{a \in {\cal L}\ |\ p \prec a\} = [p,I] \end{eqnarray}
then $(\Sigma,<,{\cal L},\prec,\wedge,\vee, \xi)=: K(\Sigma,{\cal L})$ is a state determined state property system. \end{theorem}
Proof: This proof mostly consists
of very easy verifications. We will only make the following three remarks. For all $p \in \Sigma$, $0 \not\in \xi(p)$ holds because $0 \not\in
\Sigma$. The '$\Leftarrow$' of equation (\ref{eq:xi2}) is proven as follows: $a \in
\xi(p) \Rightarrow b\in \xi(p)$ for every p in $\Sigma$ implies that $\{p \in \Sigma\ |\ p
\prec a\}
\subset \{p \in \Sigma\ |\ p \prec b\}$. Since $\Sigma$ is order-generating, this implies $a \prec b$. The state property system is state determined because $(\Sigma,<)$ is a poset.\qed 

\bigskip
\noindent
To deal with the morphisms, we will use `Galois connections'. We will quickly state the necessary results without proofs. Most of those proofs are straightforward. We will not give the results in their full generality, but will adapt them to the situation at hand. For more information we refer to (Gierz et al. 1980).

\begin{definition}{\bf (Galois connection)} Let $\cal L$ and ${\cal L}'$ be complete lattices and let $g: {\cal L} \rightarrow {\cal L}'$ and $d:{\cal L}' \rightarrow {\cal L}$ be maps. $(g,d)$ is a `Galois connection' or an `adjunction' between $\cal L$ and ${\cal L}'$ provided that
\begin{equation}
\forall\ (a,a') \in {\cal L}\times{\cal L}': a' < g(a) \Leftrightarrow d(a') < a
\end{equation}
$g$ is called the `upper adjoint' and $d$ the `lower adjoint' in $(g,d)$. $d$ is also
called a lower adjoint of $g$ and $g$ an upper adjoint of $d$. \end{definition}
In fact, adjoints determine one another uniquely: 

\begin{theorem} \label{prop:GC}
Let $\cal L$ and ${\cal L}'$ be complete lattices and let $n: {\cal L} \rightarrow {\cal L}'$ and $f:{\cal L}' \rightarrow {\cal L}$ be maps. We have: \par (1) $n$ has a (necessarily unique) lower adjoint \begin{equation}
n_{\ast}:{\cal L}' \rightarrow {\cal L}: a' \mapsto \bigwedge\{a \in {\cal L}\ |\ a' < n(a)\}
\end{equation}
(i.e.\ $n$ is an (the) upper adjoint of $n_{\ast}$) if and only if $n$ preserves infima.
\par (2) $f$ has a (necessarily unique) upper adjoint \begin{equation}
f^{\ast}:{\cal L} \rightarrow {\cal L}': a \mapsto \bigvee\{a' \in {\cal L}'\ |\ f(a') <a\}
\end{equation}
(i.e.\ $f$ is a (the) lower adjoint of $f^{\ast}$) if and only if $f$ preserves suprema.

This implies that if $f$ preserves suprema, $f^{\ast}$ exists and preserves infima, whence $(f^{\ast})_{\ast}$ exists and equals $f$. Of course the ``dual' ' holds for an
infima preserving $n$. \qed
\end{theorem}
We remark that $n: {\cal L} \rightarrow
{\cal L}'$ is said to `preserve infima' if for every family $(a_i)_i \in {\cal L}$ we have $n(\wedge_i a_i) = \wedge_in(a_i)$.

\bigskip
\noindent
We introduce morphisms of based complete lattices and show their connection to morphisms of state property systems. 

\begin{definition}{\bf (morphism of based complete lattices)} \label{def:morL0} Let $(\Sigma,{\cal L})$ and $(\Sigma',{\cal L}')$ be based complete lattices. Then a
function
$f:{\cal L} \rightarrow {\cal L}'$ is called a `morphism of based complete lattices' if
\begin{eqnarray}
f(\Sigma) &\subset& \Sigma' \\
f(\vee_ia_i)&=&\vee_if(a_i) \ \ \ \forall\ (a_i)_i \in {\cal L} \end{eqnarray}
The composition of these morphisms is given by the normal composition of functions.
\end{definition}

\begin{theorem} \label{prop:Hmor}
Consider $(m,n) \in {\bf SP}((\Sigma',{\cal L}',\xi'), (\Sigma,{\cal L},\xi))$. Then
\begin{equation}
H(m,n):= n_{\ast}:H(\Sigma',{\cal L}',\xi') \rightarrow H(\Sigma, {\cal L},\xi)
\end{equation}
is a morphism of based complete lattices. \end{theorem}
Proof: Remember that $H(\Sigma,{\cal L},\xi) = (\Sigma^{\xi},{\cal L})$ (definition
\ref{def:Hob}). We know that $n$
preserves infima (see (\ref{eq:ninf})), whence it has a suprema preserving lower adjoint
$n_{\ast}$. Next, take $s_{\xi'}(p') \in \Sigma'^{\xi'}$. We have, for $a \in {\cal L}$, $s_{\xi'}(p') < n(a) \Leftrightarrow n(a) \in \xi'(p') \Leftrightarrow a \in \xi(m(p'))$. This implies $n_{\ast}(s_{\xi'}(p')) = \bigwedge\{a \in {\cal L}\ |\ s_{\xi'}(p') < n(a)\} = \bigwedge\xi(m(p')) = s_{\xi}(m(p')) \in \Sigma^{\xi}$. \qed

\begin{theorem} \label{prop:Kmor}
Let $f:(\Sigma',{\cal L}') \rightarrow (\Sigma,{\cal L})$ be a morphism of based complete lattices. Then \begin{equation}
K(f):= (f|_{\Sigma'} ^{\Sigma},f^{\ast}) : K(\Sigma',{\cal L}') \rightarrow K(\Sigma,{\cal L})
\end{equation}
where $f|_{\Sigma'} ^{\Sigma} : \Sigma' \rightarrow \Sigma$ is the restriction to $\Sigma'$ and corestriction to $\Sigma$ of $f$ and $f^{\ast} : {\cal L} \rightarrow {\cal L}' : a \mapsto \bigvee\{a' \in {\cal L}'\ |\ f(a') <a\}$, is a morphism of state property systems.
\end{theorem}
Proof: Remember that $K(\Sigma,{\cal L}) = (\Sigma,{\cal L},\xi)$ with $\xi(p)=[p,I]$ (theorem~\ref{prop:Kob}). Take $a \in \cal L$ and $p' \in \Sigma'$. Then
$f^{\ast}(a)
\in
\xi'(p')
\Leftrightarrow p' < f^{\ast}(a) \Leftrightarrow f(p') < a \Leftrightarrow a \in
\xi(f(p')) = \xi(f|_{\Sigma'} ^{\Sigma}(p))$. \qed

\bigskip
\noindent
We will need

\begin{proposition} \label{prop:functorial} Let ${\cal L}_1,{\cal L}_2,{\cal L}_3,$ be complete lattices and let $g_1: {\cal L}_1
\rightarrow {\cal L}_2$ and
$g_2: {\cal L}_2
\rightarrow {\cal L}_3$ be two maps. If $g_1$ and $g_2$ are infima preserving then so
is $g_2 \circ g_1$ and
\begin{equation}
(g_2 \circ g_1)_{\ast} = (g_1)_{\ast} \circ (g_2)_{\ast} \end{equation}
Dually, if $g_1$ and $g_2$ are suprema preserving then so is $g_2 \circ g_1$ and
\begin{equation}
(g_2 \circ g_1)^{\ast} =g_1^{\ast} \circ g_2^{\ast} \end{equation}
\end{proposition}
Proof: We only prove the first case. For $a \in {\cal L}_3, b \in {\cal L}_1$ we have
$a<g_2g_1(b)
\Leftrightarrow (g_2)_{\ast}(a) < g_1 (b) \Leftrightarrow (g_1)_{\ast}(g_2)_{\ast}(a) <
b$. Using the uniqueness of adjoints this proves our claim. \qed

\bigskip
\noindent
Since it is quite obvious that the composition of morphisms of based complete lattices yields again such a morphism, that it is associative and that $id_{(\Sigma,{\cal L})} := id_{\cal L}$ satisfies the necessary axioms, we can safely introduce the following category. 

\begin{definition}{\bf (category of based complete lattices)} The category of based complete lattices with their morphisms is called $\bf L_{_0}$.
\end{definition}
We can now formally give the equivalence establishing functors. 

\begin{theorem} \label{prop:H}
The correspondence $H : {\bf SP} \longrightarrow {\bf L_{_0}}$ consisting of
\newline (1) the mapping
\begin{eqnarray}
|{\bf SP}| &\rightarrow& |{\bf L_{_O}}| \\ (\Sigma,{\cal L},\xi) &\mapsto& H(\Sigma, {\cal L},\xi)
\end{eqnarray}
where $H(\Sigma, {\cal L},\xi)$ is the based complete lattice $(\Sigma^{\xi},{\cal L})$ given by theorem \ref{prop:embstate}. \\ (2) for every pair of objects $(\Sigma,{\cal L},\xi),(\Sigma',{\cal L}',\xi')$
of $\bf SP$ the mapping
\begin{eqnarray}
{\bf SP}((\Sigma',{\cal L}',\xi'),(\Sigma,{\cal L},\xi)) &\rightarrow& {\bf L_{_0}}(H(\Sigma',{\cal L}',\xi'),H(\Sigma,{\cal L},\xi)) \\ (m,n) &\mapsto& H(m,n) = n_{\ast}
\end{eqnarray}
is a covariant functor.
\end{theorem}
Proof: This is, apart from some minor checks, a consequence of theorems \ref{prop:Hob}, \ref{prop:Hmor} and proposition \ref{prop:functorial}. \qed

\begin{theorem} \label{prop:K}
The correspondence $K : {\bf L_{_0}}\longrightarrow {\bf SP} $ consisting of
\newline (1) the mapping
\begin{eqnarray}
|{\bf L_{_0}}| &\rightarrow& |{\bf SP}| \\ (\Sigma,{\cal L}) &\mapsto& K(\Sigma,{\cal L})=(\Sigma,{\cal L},\xi) \end{eqnarray}
where $\xi(p) = [p,I]$ (proposition~\ref{prop:Kob}); \newline (2) for every pair of objects $(\Sigma,{\cal L}),(\Sigma',{\cal L}')$
of $\bf L_{_0}$ the mapping
\begin{eqnarray}
{\bf L_{_0}}((\Sigma',{\cal L}'),(\Sigma,{\cal L})) &\rightarrow& {\bf SP}(K(\Sigma',{\cal L}'),K(\Sigma,{\cal L})) \\ f &\mapsto& K(f) = (f|_{\Sigma'} ^{\Sigma},f^{\ast}) \end{eqnarray}
is a covariant functor.
\end{theorem}
Proof: This is, apart from some minor checks, a consequence of theorems \ref{prop:Kob}, \ref{prop:Kmor} and proposition \ref{prop:functorial}. \qed

\bigskip
\noindent
Finally, we reach
\begin{theorem}{\bf (equivalence of $\bf SP_{_0}$ and $\bf L_{_0}$)} The covariant functor $H$ restricts to the functor $H: {\bf SP_{_0}} \rightarrow {\bf L_{_0}}$ and the covariant functor $K$ corestricts to the functor $K:{\bf L_{_0}}\rightarrow{\bf SP_{_0}}$. These functors establish an equivalence of categories. Moreover,
$H\circ K = Id_{\bf L_{_0}}$.
\end{theorem}
Proof: First we remark that $K$ corestricts to the functor $K:{\bf L_{_0}}\rightarrow{\bf SP_{_0}}$ by proposition \ref{prop:Kob}.

\smallskip
\noindent
Step 1: $K \circ H$. Consider $(\Sigma,{\cal L},\xi) \in |{\bf SP_{_0}}|$. Then
$KH(\Sigma,{\cal L},\xi) = (\Sigma^{\xi},{\cal L},\overline{\xi})$ with
\begin{eqnarray}
\overline{\xi}:\Sigma^{\xi} &\rightarrow& {\cal P}({\cal L}) \\ a_p = s_{\xi}(p) &\mapsto& [a_p,I]
\end{eqnarray}
Also, if $(m,n) \in {\bf SP_{_0}}((\Sigma',{\cal L}',\xi'),(\Sigma, {\cal L},\xi))$ then
$KH(m,n) = K(n_{\ast}) = (n_{\ast}|_{\Sigma'^{\xi'}} ^{\Sigma^{\xi}}, n)$.

\smallskip
\noindent
Step 2: $Id_{\bf SP_{_0}}\cong KH$. For $(\Sigma,{\cal L},\xi) \in |{\bf SP_{_0}}|$ define
\begin{eqnarray}
&&\eta_{(\Sigma,{\cal L},\xi)}:(\Sigma,{\cal L},\xi) \rightarrow KH(\Sigma,{\cal L},\xi)\\
&&\eta_{(\Sigma,{\cal L},\xi)} = (s_{\xi},id_{_{\cal L}}) \end{eqnarray}
Then $\eta = (\eta_{(\Sigma,{\cal L},\xi)}) : Id_{\bf SP_{_0}} \rightarrow KH$ is a natural
isomorphism. First we verify that $\eta_{(\Sigma,{\cal L},\xi)}$ is a morphism of
$\bf SP_{_0}$. Indeed, for $a\in {\cal L}$ and $p \in {\Sigma}$ we have $a \in \overline{\xi}(s_{\xi}(p)) \Leftrightarrow s_{\xi}(p) < a \Leftrightarrow id_{_{\cal L}}(a) = a \in \xi(p)$. To show that $(s_{\xi},id_{_{\cal L}})$ is an isomorphism, we only have to prove that
$s_{\xi}:\Sigma \rightarrow \Sigma^{\xi}$ is bijective (proposition~\ref{prop:statpropiso}) and this follows from $\Sigma^{\xi} = s_{\xi}(\Sigma)$ and proposition \ref{prop:embstate}. Finally we prove the naturality of $\eta$. Take $(m,n) \in {\bf SP_{_0}}((\Sigma',{\cal L}',\xi'),(\Sigma,{\cal L},\xi))$. We have to show: $n_{\ast} \circ s_{\xi'} = s_{\xi}\circ m$. This has been done in the proof of theorem~\ref{prop:Hmor}. 

\smallskip
\noindent
Step 3: $HK = Id_{\bf L_{_0}}$. Let $f$ be a morphism of $\bf L_{_0}$. We have
\begin{equation}
f \stackrel{K}{\longmapsto} (f|_{\Sigma'} ^{\Sigma},f^{\ast}) \stackrel{H}{\longmapsto}
(f^{\ast})_{\ast} = f
\end{equation}
Next, consider a based complete lattice $(\Sigma,{\cal L})$. Then $K(\Sigma,{\cal L}) =
(\Sigma,{\cal L},\xi)$ with $\xi(p) = [p,I]$ for $p\in \Sigma$. This implies that
$s_{\xi}(p) = \bigwedge \xi(p) = p$, whence $\Sigma^{\xi} = \Sigma$. Therefore $HK(\Sigma,{\cal L}) = H(\Sigma,{\cal L},\xi) = (\Sigma^{\xi},{\cal L}) = (\Sigma,{\cal
L})$. \qed

\begin{theorem}
We have the following equivalences of categories: \begin{eqnarray}
{\bf Cls} &\approx& {\bf SP} \\
{\bf Cls_{_0}} &\approx& {\bf SP_{_0}}\ \approx\ {\bf L_{_0}} \end{eqnarray}
\end{theorem}
This last theorem shows that a state determined entity can ``equivalently'' be described by a T$_{_0}$ closure space (where the states are the points ---or the point closures--- and the properties are represented by the closed subsets), a state determined state property system or a based complete lattice (where the
states form an order-generating subset of the property lattice). 

In (Ern\'{e} 1984) is shown the direct equivalence between $\bf Cls_{_0}$ and $\bf L_{_0}$.

\section{Construction of the (co)product of two state property systems} \label{sec:tens}

We will now construct the
product of two state property systems in $\bf SP$. For the necessary category theory we refer to (Borceux 1994).

\begin{theorem}
Let $(\Sigma_1,{\cal L}_1,\xi_1)$ and $(\Sigma_2, {\cal L}_2,\xi_2)$ be state property systems (objects of $\bf SP$). Then $(P,(s_1,s_2))$ is the (up to isomorphism (see Borceux 1994 , proposition 2.2.2)) product of $(\Sigma_1,{\cal L}_1,\xi_1)$ and $(\Sigma_2, {\cal L}_2, \xi_2)$ in $\bf SP$, where $P$ is the state property system $(\Sigma,<,{\cal L},\xi)$ with
\begin{eqnarray}
\Sigma &=& \Sigma_1 \times \Sigma_2 \\
(p_1,p_2)< (q_1,q_2) &\Leftrightarrow& p_1 < q_1 \mbox{ and } p_2 < q_2 \mbox{ for }p_i,q_i \in \Sigma_i
\end{eqnarray}
\begin{eqnarray}
{\cal L} &=& {\cal L}_1 \coprod {\cal L}_2\\ &=& \{(a_1,a_2)\ |\ a_1 \in {\cal
L}_1, a_2
\in {\cal L}_2, a_1 \neq 0_1, a_2 \neq 0_2 \} \bigcup \{0\} \end{eqnarray}
equipped with the following partial order relation: \begin{eqnarray}(a_1,a_2) &<& (b_1,b_2) \Leftrightarrow a_1 < b_1 \mbox{ and } a_2 < b_2 \\ 0 &<& (a_1,a_2) \mbox{ for all } (a_1,a_2) \end{eqnarray} and lattice operations:
\begin{eqnarray}\bigwedge_i(a_{1}^{i}, a_{2}^i) &=& \left\{ \begin{array}{ll} (\wedge_i a_{1}^i,
\wedge_i a_{2}^i) & \mbox{if
$\wedge_i a_{1}^i
\neq 0_1$ and $\wedge_i a_{2}^i \neq 0_2$} \\ 0 & \mbox{otherwise} \end{array} \right. \\
\bigvee_i(a_{1}^{i}, a_{2}^i) &=& (\vee_i a_{1}^i, \vee_i a_{2}^i). \end{eqnarray}
and with
\begin{eqnarray}
\xi: \Sigma &\rightarrow& {\cal P}({\cal L})\\ (p_1,p_2) &\mapsto& \{(a_1,a_2) \in {\cal L}\ |\ a_1 \in \xi_1(p_1), a_2 \in \xi_2(p_2)\}
\end{eqnarray}
and $s_i = (\pi_i,\imath_i)$ with
\begin{equation}
\pi_i:\Sigma \rightarrow \Sigma_i:(p_1,p_2) \mapsto p_i \end{equation}
\begin{eqnarray}
\imath_1: {\cal L}_1 &\rightarrow& {\cal L}_1 \coprod {\cal L}_2\\ 
a_1 &\mapsto& (a_1,I_2) \ \mbox{ if }a_1\not=0_1\\ 0_1 &\mapsto& 0
\end{eqnarray}
\begin{eqnarray}
\imath_2: {\cal L}_2 &\rightarrow& {\cal L}_1 \coprod {\cal L}_2\\ 
a_2 &\mapsto& (I_1,a_2) \ \mbox{ if }a_2\not=0_2\\ 0_2 &\mapsto& 0
\end{eqnarray}
\end{theorem}
Proof:

\smallskip
\noindent
Step 1: $P \in |{\bf SP}|$. We have to check the conditions of definition~\ref{def:statprop}. After noting that $\xi(p_1,p_2)= \xi_1(p_1) \times\xi_2(p_2)$,
$I=(I_1,I_2)$ this requires more writing than thinking. 

\smallskip
\noindent
Step 2: $s_i$ is a morphism of $\bf SP$. We check equation (\ref{eq:covar1}). Let
$(p_1,p_2) \in \Sigma$,
$a_1\in {\cal L}_1$ and take $a_1 \not=0_1$ (the other case is trivial). Then
$\imath_1(a_1) = (a_1,I_2) \in \xi(p_1,p_2) \Leftrightarrow a_1 \in \xi_1(p_1)=\xi_1(\pi_1(p_1,p_2))$ ($I_2 \in \xi_2(p_2)$ always holds). 

\smallskip
\noindent
Step 3. Let $Q=(\Sigma',{\cal L}',\xi')$ be a state property system and consider two morphisms of $\bf SP$:
$(m_1,n_1):Q \rightarrow (\Sigma_1,{\cal L}_1,\xi_1)$ and $(m_2,n_2):Q \rightarrow (\Sigma_2,{\cal L}_2,\xi_2)$. We define $(m,n)$ by
\begin{eqnarray}
m:\Sigma' \rightarrow \Sigma:p'&\mapsto& (m_1(p'),m_2(p')) \\ \nonumber \\
n:{\cal L}\rightarrow{\cal L}':(a_1,a_2) &\mapsto& n_1(a_1) \wedge n_2(a_2)\\ 0&\mapsto&0'
\end{eqnarray}
Then $(m,n):Q \rightarrow P$ is a morphism of $\bf SP$. Indeed for $a_i \in {\cal L}_i$,
$a_i\not=0_i$,$i=1,2$ (the zero case is trivial) and $p'\in \Sigma'$ we have $(a_1,a_2) \in
\xi(m(p'))=\xi(m_1(p'),m_2(p'))
\Leftrightarrow a_1 \in \xi_1(m_1(p')), a_2\in \xi_2(m_2(p')) \Leftrightarrow n_1(a_1)\in
\xi'(p'), n_2(a_2)\in \xi'(p') \Leftrightarrow n(a_1,a_2)=n_1(a_1) \wedge n_2(a_2) \in
\xi'(p')$.

\smallskip
\noindent
Step 4: $s_i \circ (m,n) = (m_i,n_i)$. We have to show $\pi_i \circ m=m_i$ and $n \circ
\imath_i = n_i$. The first is trivial. The second isn't difficult either: for
$a_1 \not= 0_1$ (other case again trivial) we have $n(\imath_1(a_1)) = n(a_1,I_2) =
n_1(a_1)\wedge n_2(I_2) = n_1(a_1)$ since $n_2(I_2) = I$.

\smallskip
\noindent
Step 5: We have to show that $(m,n)$ is the only morphism such that $(m_i,n_i)=s_i \circ (m,n)$. Clearly $m$ is the only function such that $m_i=\pi_i \circ m$. $n_i = n\circ \imath_i$ clearly implies that for $a_i\in {\cal L}_i$, $a_i\not=0$ ($n(0)=0'$ must hold because $n$ should be a morphism) we have $n(a_1,a_2)= n((a_1,I_2) \wedge (I_1,a_2))= n(a_1,I_2)\wedge n(I_1,a_2) = n(\imath_1(a_1))\wedge n(\imath_2(a_2))=n_1(a_1)\wedge n_2(a_2)$. \qed

\bigskip
\noindent
We make some remarks. (1) If we consider the opposite category $\bf SP^{\rm op}$,
this product becomes a coproduct. This is a generalization of the coproduct (tensor product) of property lattices of (Aerts 1984a), which is in fact a product
in $\bf L_{_0}$ (or a coproduct in $\bf L_{_0}^{\rm op})$. (2) As the finite coproduct of (Aerts 1984a) has been generalized to arbitrary coproducts (Aerts and
Valckenborgh 1998), the product of the previous theorem can also be constructed for
arbitrary families of state property systems. (3) Even before we did the calculations for the previous theorem, we knew the category $\bf SP$ had arbitrary
products, since it is equivalent with the topological (and hence complete) category
$\bf Cls$.

\section{References}

Aerts, D., 1981, {\it The one and the many}, Doctoral Thesis, Brussels Free University.

\bigskip
\noindent
Aerts, D., 1982, ``Description of many physical entities without the paradoxes encountered in quantum mechanics", {\it Found. Phys.\/}, {\bf 12}, 1131.

\bigskip
\noindent
Aerts, D., 1983a, ``Classical theories and Non Classical Theories as a Special Case of a More General Theory", {\it J. Math. Phys.\/} {\bf 24}, 2441.

\bigskip
\noindent
Aerts, D., 1983b, ``The description of one and many physical systems", in {\it Foundations of Quantum Mechanics\/}, eds. C. Gruber, A.V.C.P. Lausanne, 63.

\bigskip
\noindent
Aerts, D., 1984a, ``Construction of the tensor product for the lattices of properties of physical entities", J. Math. Phys. {\bf 25}, 1434.

\smallskip \noindent Aerts, D., 1984b, ``The missing elements of reality in the description of quantum mechanics of the EPR paradox situation", Helv. Phys. Acta {\bf 57}, 421.

\bigskip
\noindent Aerts D., 1985, ``The physical origin of the EPR paradox and how to violate Bell inequalities by macroscopic systems", in {\it Symposium on the Foundations of Modern
Physics,} P. Mittelstaedt, and P. Lahti, eds. World Scientific, Singapore.

\bigskip
\noindent
Aerts, D., 1994, ``Quantum Structures, Separated Physical Entities and Probability", {\it Found. Phys.\/} {\bf 24}, 1227.

\bigskip
\noindent
Aerts, D., 1998, ``Foundations of quantum physics: a general realistic and operational approach", to appear in International Journal of Theoretical Physics.

\bigskip
\noindent
Aerts, D., Coecke, B., Durt, T. and Valckenborgh, F., 1997a, ``Quantum, Classical and Intermediate; a Model on the Poincar\'e Sphere, {\it Tatra Mountains Math. Publ.\/} {\bf 10}, 225. 

\bigskip
\noindent
Aerts, D., Coecke, B., Durt, T. and Valckenborgh, F., 1997b, ``Quantum, Classical and Intermediate; the Vanishing Vector Space Structure", {\it Tatra Mountains Math. Publ.\/} {\bf 10}, 241. 

\bigskip \noindent Aerts, D. and Valckenborgh, F., 1998, ``Lattice extensions and the description of compound entities", FUND, Brussels Free University, preprint.

\bigskip
\noindent
Borceux F., 1994, {\em Handbook of categorical algebra I}, Encyclopedia of mathematics and its applications, Cambridge University Press.

\bigskip
\noindent
Cattaneo G., dalla
Pozza C., Garola C. and Nistico G., 1988, ``On the logical foundations of the Jauch-Piron approach to quantum physics", Int. J. Theor. Phys. {\bf 27}, 1313.

\bigskip
\noindent
Cattaneo G. and Nistico G., 1991, ``Axiomatic foundations of quantum physics: Critiques and misunderstandings. Piron's question-proposition system", Int. J. Theor. Phys., {\bf 30} , 1293. 

\bigskip
\noindent
Cattaneo G. and Nistico G., 1992, ``Physical content of preparation-question structures and Bruwer-Zadeh lattices", Int. J. Theor. Phys. {\bf 31}, 1873. 

\bigskip
\noindent
Cattaneo G. and Nistico G., 1993, ``A model of Piron's preparation-question structures in Ludwig's selection structures", Int. J. Theor. Phys., {\bf 32}, 407.

\bigskip
\noindent Daniel, W., 1982, ``On non-unitary evolution of quantum systems", Helv. Phys. Acta, {\bf 55}, 330.

\bigskip
\noindent d'Emma G., 1980, ``On quantization of the electromagnetic field", Helv. Phys. Acta, {\bf 53}, 535. 

\bigskip
\noindent
Ern\'{e} M., 1984, {\it Lattice representations for categories of closure spaces}, in ``Categorical Topology'', Proc.\ Conf.\ Toledo Ohio 1983, 197--222, Heldermann Verlag, Berlin. 

\bigskip
\noindent
Foulis D., Piron C. and Randall C., 1983, ``Realism, operationalism and quantum mechanics", Found. Phys. {\bf 13}, 843. 

\bigskip
\noindent Foulis D. and Randall C., 1984, ``A note on the misunderstanding of Piron's axioms for quantum mechanics", Found. Phys., {\bf 14}, 65.

\bigskip
\noindent
Gierz G., Hofmann K.H., Keimel K., Lawson J.D., Mislove M. and Scott D.S., 1980, {\it A compendium of continuous lattices}, Springer-Verlag, Berlin - Heidelberg - New York, 1980.

\bigskip
\noindent Giovannini N. and Piron C., 1979, ``On the group-theoretical foundations of
classical and quantum physics: Kinematics and state spaces", Helv. Phys. Acta, {\bf 52}, 518.

\bigskip
\noindent Gisin, N., 1981, ``Spin relaxation and dissipative Schr\"odinger like evolution equations", Helv. Phys. Acta {\bf 54}, 457.

\bigskip
\noindent Jauch J. M. and Piron C., 1965, ``Generalized localisability", Helv. Phys. Acta, {\bf 38}, 104. 

\bigskip
\noindent Ludwig G. and Neumann H., 1981, ``Connections between different approaches to the foundations of quantum mechanics", in {\it Interpretation and Foundations of Quantum Mechanics,} H. Neumann,su eds. B. I. Wissenschafts-Verlag, Bibliographisches Institut, Mannheim.

\bigskip
\noindent
Moore, D., 1995, ``Categories of Representations of Physical Systems", Helv. Phys.
Acta, {\bf 68}, 658.

\bigskip
\noindent
Piron, C., 1964, ``Axiomatique Quantique", Helv. Phys. Acta, {\bf 37}, 439.

\smallskip
\noindent Piron, C., 1969, ``Les r\'egles de superselection continues", Helv. Phys. Acta {\bf 43}, 330.

\bigskip
\noindent
Piron, C., 1976, {\it Foundations of Quantum Physics,\/} Reading, Mass., W. A. Benjamin.

\bigskip
\noindent Piron, C., 1989, ``Recent Developments in Quantum Mechanics", Helv. Phys. Acta, {\bf 62}, 82.

\bigskip
\noindent
Piron, C., 1990, {\it M\`ecanique Quantique: bases et applications,}, Press Polytechnique de Lausanne.

\bigskip
\noindent
Randall C. and
Foulis D., 1983, ``Properties
and operational propositions in quantum mechanics", Found. Phys. {\bf 13}, 843.

\end{document}